# Acceptability of AI Assistants for Privacy: Perceptions of Experts and Users on Personalized Privacy Assistants

**Meihe Xu** [1] *  **Aurelia Tamò-Larrieux** [2], **Arianna Rossi**[3]

[1]Maastricht University  [2]University of Lausanne  [3]Sant'Anna University of Advanced Studies

**Abstract**

Individuals increasingly face an overwhelming number of tasks and decisions. To cope with the new reality, there is growing research interest in developing intelligent agents that can effectively assist people across various aspects of daily life in a tailored manner, with privacy emerging as a particular area of application. Artificial intelligence (AI) assistants for privacy, such as personalized privacy assistants (PPAs), have the potential to automatically execute privacy decisions based on users' pre-defined privacy preferences, sparing them the mental effort and time usually spent on each privacy decision. This helps ensure that, even when users feel overwhelmed or resigned about privacy, the decisions made by PPAs still align with their true preferences and best interests. While research has explored possible designs of such agents, user and expert perspectives on the acceptability of such AI-driven solutions remain largely unexplored. In this study, we conducted five focus groups with domain experts (n = 11) and potential users (n = 26) to uncover key themes shaping the acceptance of PPAs. Factors influencing the acceptability of AI assistants for privacy include design elements (such as information sources used by the agent), external conditions (such as regulation and literacy education), and systemic conditions (e.g., public or market providers and the need to avoid monopoly) to PPAs. These findings provide theoretical extensions to technology acceptance models measuring PPAs, insights on design, and policy implications for PPAs, as well as broader implications for the design of AI assistants.

## Introduction

We have seen a rapid increase in research and development of artificial intelligence (AI) assistants (Alowais et al. 2023; Maedche et al. 2019; Mariani, Hashemi, and Wirtz 2023). AI assistants are technologies designed to assist individuals who want to accomplish a specific task (Maedche et al. 2019). The applications for AI assistants seem limitless, such as AI doctors that can make diagnoses to aid human doctors in patient treatment (Duffourc 2023), AI personal trainers who are able to evaluate a person's training performance and offer feedback (Chariar et al. 2023; Novatchkov and Baca 2013), AI tutors who are able to personalize study content and quizzes based on a student's existing level of knowledge and progress (Baillifard et al. 2025; Kim and Kim 2020), and AI assistants that can assist an individual with their privacy management (Morel, Iwaya, and Fischer-Hübner 2025).

Among these varied applications of AI assistants, the latter branch concerned with helping users with privacy management entails personalized privacy assistants (PPAs), defined as *"intelligent agents capable of learning the privacy preferences of their users over time, semi-automatically configuring many settings, and making many privacy decisions on their behalf"* (Privacy Assistant Project, accessed in 2025). Under the defining characteristics by Maedche et al. (2019), PPAs belong to the family of AI assistants, with users able to delegate tasks of privacy management to a technology. The current practice of user privacy management neglects the nuanced and complex nature of individual privacy preferences (e.g., context, see Nissenbaum 2011) and leaves little room for negotiation between service providers and users (i.e., 'take it or leave it' approach). People are expected to read disclosures like privacy policies before they make a decision to consent or reject the processing of their data. However, such disclosures are often long, jargon-laden, and incomprehensible to an average user (Ibdah et al. 2021; Fabian, Ermakova, and Lentz 2017; Wagner 2023; Vu et al. 2007). Often, users opt to skip reading privacy disclosures (Obar and Oeldorf-Hirsch 2020; Steinfeld 2016), effectively making their subsequent consent decisions less meaningful. On the other hand, although the binary choices (accept/reject) may seem to make decision-making easier for users, the sheer number of decisions still burdens users due to cognitive overload (Bawden and Robinson 2009) or bounded rationality (Acquisti and Grosslags 2005), and could eventually lead to user privacy cynicism and resignation from privacy management (Acquisti, Brandimarte, and Loewenstein 2015; Draper 2017; Hoffmann, Lutz, and Ranzini 2016). If users could delegate decision-making to PPAs tailored to their individual preferences or dispositions, this technology might effectively assist users in managing their privacy.

Researchers designing PPAs propose different and non-exclusive functions, such as personalized notifications (Das et al. 2018; Tran et al. 2023), personalized recommendations (Chang and Barber 2023; Liu et al. 2016), or automated decision-making (Ayci et al. 2023). Morel, Iwaya, and Fischer-Hübner (2025) in their systematization of knowledge study on AI-driven PPAs further categorize AI-driven PPAs by types of decisions, types of AI used, contexts, ar-chitecture, and user control over PPA decisions. Despite this growing attention, one integral stakeholder is underrepresented in the discussion of AI assistants: the users. While Morel, Iwaya, and Fischer-Hübner (2025) emphasize the need for user studies to validate PPAs, we argue that an essential preceding step is understanding user acceptance of PPAs, considering that the ready availability and possibility of PPAs are not without caution or concerns.

---

*Corresponding author: m.xu@maastrichtuniversity.nl

Ethical issues related to AI assistants have been raised in the literature. For example, Danaher (2018) discussed the concerns of AI assistants contributing to cognitive degeneration, threatening individual autonomy, and reducing authenticity in personal communications, when users delegate tasks to AI systems (Danaher 2018). Gabriel et al. (2024) provided a comprehensive coverage of the ethical and societal risks that could be posed by AI assistants, such as concerns about value alignments or the risks of misinformation. AI assistants that deal with privacy could add another layer of concerns, specifically connected to their handling of privacy decisions. For instance, when individuals were asked to provide feedback on PPA designs in an Internet of Things (IoT) context, some opposed PPAs providing personalized recommendations for fear of losing control over privacy decisions (Colnago et al. 2020). Stöver et al. (2023) in their pilot interviews uncover individuals' concerns about personal privacy assistants such as the meddling of PPAs on one's privacy. These concerns bring to light that user perspectives are needed, as a promising technology does not automatically ensure user acceptance. Before making PPAs available to the general public, it is necessary to understand how acceptable they are to users and what conditions must be satisfied to make them acceptable. In this article, we center our focus on the perspectives of users, comparing them further with those of experts. Therefore, we aim to explore *"what are the key themes influencing the acceptability of personalized privacy assistants (PPAs) perceived by users and experts?"*

By answering our research question, we make four contributions:

- We provide novel insights into both user and expert perspectives on PPA acceptability and disentangle design, external, and systemic factors that lead to greater acceptability of such assistants.
- We propose possible theoretical extensions for existing acceptance models on AI assistants for privacy management based on our qualitative analysis.
- We shed light on the design and regulatory implications of AI assistants within the domain of privacy.
- We reflect on PPAs' frictionless design, their potential contribution to the instrumentarian power of dominant commercial providers, and whether they suffice to address the privacy issues users have.

In the following sections, we first describe how the acceptability of AI assistants and PPAs in particular has been studied so far. Upon this basis, we elaborate on our methodological approach, including the use of focus groups and thematic analysis. We then present the three principal themes and their corresponding sub-themes in the Results section.

Drawing on insights from both users and experts, we discuss how themes from the focus groups could contribute to existing technology acceptance models for PPA acceptance. We further discuss the implications for diverse stakeholders and lessons for AI assistants in general, and reflect on PPAs as a frictionless technology.

## Studying the Acceptability of AI Assistants for Privacy Management

In this article, we use "acceptance" and "acceptability" interchangeably. We acknowledge that the two words have nuanced differences. Whereas "acceptability" pertains to the evaluation of technology before its use, focusing on the perceptions of a technology that has yet to be employed; "acceptance", on the other hand, relates to the assessment of a technology after its use (Koelle et al. 2019; Nadal, Doherty, and Sas 2019). In this study, both words refer to the definition of acceptability, recognizing that PPAs are still an emerging technology without fully developed prototypes for users to interact with. Jiang et al. (2024) demonstrates that studies on AI application acceptance typically categorize acceptance into three key dimensions: behavior, behavioral intention, and perception. Although various theoretical perspectives exist, the most widely employed frameworks remain the Technology Acceptance Model (TAM), its extensions, and the Unified Theory of Acceptance and Use of Technology (UTAUT) (Jiang et al. 2024; Kelly, Kaye, and Oviedo-Trespalacios 2023). Research on AI assistants often relies on these frameworks, adjusting them with new constructs or proposing contextual variables (Balakrishnan et al. 2024; Choung, David, and Ross 2023; Pan et al. 2024; Xiong et al. 2024).

Quantitative methods with established theoretical models are not always the most suitable approach for understanding acceptance. According to Vogelsang, Steinhüser, and Hoppe (2013), quantitative approaches are effective primarily for testing existing theories, whereas qualitative methods are better suited for developing new theories or discovering new constructs to existing theories. For instance, Hasija and Esper (2022) combined qualitative insights obtained from textual analyses and interviews to extend the UTAUT model for the use of AI in supply chain management. In addition to studies on technology acceptance, qualitative methods such as semi-structured interviews and focus groups are often used to reveal users', experts', and other stakeholders' understandings and perceptions of artificial intelligence (see, for example, Chedrawi, Kazoun, and Kokkinaki 2024; Morgenstern et al. 2021; Troshani et al. 2021). Given that PPAs are still relatively emerging, qualitative methodologies can offer unique advantages in uncovering factors that would influence user acceptance of PPAs.

## Methodology

In this article, we opted for focus groups due to their ability to "obtain data regarding the ideas, attitudes, understanding and perceptions, as well as learn the typical vocabulary and thinking patterns of the selected sample from the target population when they talk about a particular topic" (Plummer-

D'Amato 2008, p. 69). Five focus groups were conducted (see Table 1 for details). Out of the five focus groups, the first one (FG 1) was an expert focus group consisting of legal scholars, social scientists, and one industry insider involved in the development of PPAs. The expert focus group was part of a workshop organized by the authors on PPAs. The remaining four (FG 2–5) were user focus groups, with participants recruited from a law faculty of a Dutch university using snowball sampling. These user focus groups were further divided based on participants' familiarity with data protection law—some had taken a data protection law course, while others did not. Participants were not compensated with money, but were provided with refreshments (snacks and drinks) during and after their focus groups. Participants could withdraw from the study at any time without consequences. The study was approved by the ethical review board of the lead author's institution. Prospective participants were given an information letter about the study at least 24 hours prior to their focus group, and consents were acquired before the start of each focus group. The first focus group (FG 1) took approximately 90 minutes while the rest (FG 2-5) around 60 minutes.

| Focus group | Composition of participants | Date held | No. participants |
| --- | --- | --- | --- |
| 1 | Experts | May 2023 | 12* |
| 2 | Users without DPLK** | April 2024 | 8 |
| 3 | Users with DPLK | April 2024 | 6 |
| 4 | Users without DPLK | April 2024 | 7 |
| 5 | Users with DPLK | April 2024 | 5 |

Table 1: Details of focus groups

*By discipline: Law 8, Social Science 2, Philosophy 1, Industry 1
**DPLK = Data protection law knowledge

The number of participants per focus group in this study ranged from 5 to 11. In the literature, there is a lack of consensus on the ideal size per focus group, with recommendations from as few as three (Lane et al. 2001), six to eight (Krueger and Casey 2002), to eight to twelve (Stewart and Shamdasani 2014). Instead of numbers, some researchers place an emphasis on ensuring meaningful interactions among participants in a focus group (Plummer-D'Amato 2008). Group sizes in this study align with these recommendations.

Prior to the discussion of each focus group, participants were introduced to a hypothetical PPA named *TamagotchIA*, which was inspired by the speculative work of the French Data Protection Authority (CNIL) (Courmont et al. 2021).

The described capabilities of TamagotchIA stated that the assistant uses a hybrid approach to personalize its actions, integrating both manual and automated learning mechanisms through its interactions with users. It leverages declarative knowledge (explicit user preferences) as well as observational knowledge (inferred behaviors and contextual cues).

The focus group discussions were facilitated by question guides (see Appendix A): The first guide, developed for the expert focus group, focused on mandatory, desirable, and inadmissible features of PPAs and the rationale behind them. Insights from this discussion informed the development of the second and third question guides. The second guide, used for FG 2-3, explored the providers of PPAs, mandatory features, sources of personalization, and participants' acceptance or rejection of PPAs, along with their rationales. The third guide, used for FG 4-5, was structured around UTAUT2, with its constructs translated into questions to elicit user responses on PPA acceptance. However, it is important to clarify that the questions were designed to encourage user responses and interactions and were not used as a theoretical framework for analysis. Combined, the iterative question guide design and the flexibility in question phrasing and focus afford the authors a deeper and broader exploration regarding the acceptability of PPAs by experts and users.

The discussions held within the focus groups were recorded, transcribed, and anonymized. Transcripts were analyzed using reflexive thematic analysis by the first author using Atlas ti, during which coding was "open and organic, with no use of any coding framework" (Braun and Clarke 2021b, p.334). We opted for reflexive thematic analysis because it allowed us to engage with the data through an iterative back-and-forth between the transcript, codes, and themes, allowing us to organically identify explicit or latent themes related to PPA acceptability (Braun and Clarke 2019, 2021b,a). Following Braun and Clarke's six phases (Braun and Clarke 2006), the first author immersed in the data, coded it, identified and refined themes, and named them, with the findings reported in the section below.

## Results

Three main themes were identified from the focus groups that impact the acceptability of PPAs: (1) Design Elements; (2) External Conditions; (3) Systemic Conditions. Figure 1 shows the themes and sub-themes. Although we identified three distinct themes, they interact with each other, jointly shaping the acceptability of PPAs among users and experts. For example, the degree of transparency within Design Elements is likely influenced by Regulation in External Conditions and by whether a PPA is offered as a public service or as a market product in Systemic Conditions. Throughout the Results section, the collective term "participants" is used whenever the perspectives of the expert and user groups converge; only where their views diverge do we explicitly attribute remarks to experts or users.

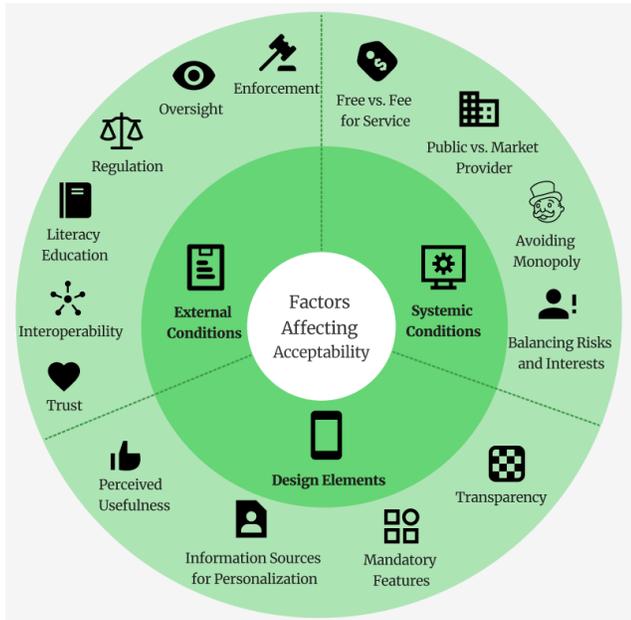

Figure 1: Themes and sub-themes affecting assistant acceptability

### Design elements

**Perceived usefulness:** Participants viewed PPAs as a more time-efficient alternative to "privacy self-management", allowing users to avoid spending precious time reading privacy disclosures or making cookie decisions, while enabling users to make decisions on a more granular level than humans and without the susceptibility to manipulations. For example, PPAs could help ensure that their consent decisions align with their actual preferences rather than accepting all requests. Thus, PPAs were considered a better alternative compared to leaving users to make decisions on their own.

Even if participants acknowledged that the initial setup of privacy preferences might require additional time, they agreed that, in the long run, PPAs would still be more time-saving. In addition, PPAs were also perceived as a means to address shortcomings in human privacy decision-making. Participants noted that PPAs could help them better understand the terms of their consent and enhance their knowledge regarding the processing of their personal information, such as who has access to their data.

**Information sources for personalization:** Participants generally concurred that a PPA should rely on users' pre-defined privacy preferences as its primary source for personalization. The questions used to set up user privacy preferences need to be constructed in a way that accurately captures users' true privacy preferences. This is particularly important to prevent situations where individuals with limited privacy literacy or knowledge misinterpret the questions or provide responses that do not reflect their actual preferences. With respect to more automatic approaches to extract user preferences, such as based on behavioral data, demographics, or other personal information, opinions on their acceptability diverged. While some participants supported the use of privacy-related behavioral data to enhance PPAs' effectiveness, others expressed serious concerns about the privacy intrusion such usage entails.

**Transparency:** Participants emphasized the need for transparency in PPAs' data processing, such as the need for a PPA to disclose its data processors and other entities with access to users' personal information. Participants further believed that a PPA must provide clear explanations of its decision-making process and actions to users. It should be able to articulate how it interprets privacy statements or requests from third-party companies and how it makes decisions based on a user's pre-defined preferences. A PPA should explain why it denies certain requests and, in cases where companies with similar privacy statements receive a different response, clarify the reasoning behind those differing decisions.

**Mandatory features:** In the expert focus group, mandatory features revolved around the execution of user rights (consent and opt-out), respecting user control, and ensuring cross-platform compatibility. Expert participants emphasized that users should have the right to opt out of PPA data collection and processing while ideally still being able to use the service. One expert participant viewed opting out as the ability to refuse the delegation of privacy decision-making to a PPA altogether. They warned that, if PPAs become ubiquitous, people could become so reliant on them that forgoing their use would no longer be a realistic choice.

In the user focus groups, participants emphasized the importance of a kill switch function to quickly and easily execute users' deletion requests, the importance of PPA explanation to users, and of users retaining control. Participants stated that PPAs must provide real-time updates or periodic reports to their users. Such updates should entail decisions made by the PPAs, the reasoning behind the decisions, and their consequences (e.g., what type of information was shared and with whom after a PPA decision was made). To retain user control, when a PPA applies default settings based on a user's privacy preferences, users should have the ability to review and modify these settings as needed. PPAs should not make decisions automatically without user approval, and their decisions could be overridden by users.

### External conditions

**Trust:** Trust is closely intertwined with transparency. On the one hand, transparency of a PPA can foster user trust, increasing its acceptability. For instance, allowing community validation through open-source accessibility would reinforce trust in using that PPA. On the other hand, if transparency is absent, such as when companies cite trade secrets or intellectual property rights as reasons for not disclosing their code, users may have no choice but to rely on blind trust in the company. In such cases, trust in the PPA becomes dependent on the perceived trustworthiness of the entity or individuals behind the technology rather than the transparency of the PPA.

**Literacy education:** Participants stressed the need for privacy literacy to ensure that individuals use PPAs responsibly to better know their privacy preferences and how decisions made by PPAs reflect their preferences. Beyond general privacy literacy, participants also stressed the need for education related explicitly to PPAs. Participants were concerned about users' lack of knowledge regarding how a PPA uses their personal data, how it operates, and the extent of their consented permission. They suggested that clear instructions should be provided before users set their privacy preferences to ensure they fully understand the risks, such as the implications of accepting all cookies. To accommodate users with varying levels of privacy literacy, participants argued that privacy preference questions should be designed in a way that aligns with the user's level of literacy.

**Interoperability:** The importance of establishing interoperability standards to support the functioning of PPAs was mentioned across focus groups. It was considered important due to the current absence of interoperable standards on consent management, meaning that a PPA needs to reverse engineer technical solutions to each company's cookie consent framework in order to work. Furthermore, PPAs should be compatible with various digital platforms and be able to operate across multiple devices. PPAs that only work on one device or platform would not be considered acceptable due to their loss in the trade-off, namely, providing personal data for tailored privacy protection that only operates on a limited number of platforms or devices.

**Regulations:** PPAs must comply with existing laws and regulatory frameworks. And, if necessary, appropriate regulations should be established. The experts emphasized the need for clarity on accountability in cases where PPAs' decisions result in harm to users.

One user expressed the concern that big tech companies might invoke trade secret protections as a strategy to oppose or hinder the adoption of PPAs, for example, by resisting PPAs to connect to their consent management infrastructure. Others in the user focus groups with knowledge of data protection laws highlighted concerns regarding PPA compliance with EU regulations and the potential need for additional legal regulations. They questioned the legality of PPAs consenting on behalf of users, arguing that such delegation may not constitute valid consent under GDPR requirements. Participants also noted that if PPAs rely on biometric data for personalization, they could be classified as high-risk systems under the AI Act. Resonating with the experts, they emphasized the need for mechanisms that allow users to challenge or take legal action against PPA developers or the software itself in cases where PPAs fail to function as expected or cause harm to users.

**Oversight:** Oversight involves multiple stakeholders, including the PPA community, users, and government oversight bodies. While the expert discussions focused primarily on the oversight of PPAs' code (with their ideal of open-source models) and decision-making processes, users focused more on EU regulatory bodies and the role these institutions need to play to ensure PPAs work in their best interest. One user suggested that authorities should ensure that users have access to logs of PPAs to track and verify whether decisions align with their privacy preferences. Additionally, participants proposed the use of an external agent to assist users in evaluating whether a PPA functions correctly. Such evaluations were also mentioned by the experts who discussed the oversight duties of Data Protection Authorities (DPAs). In cases where complaints are filed, DPAs could investigate and audit PPA systems, including reviewing the source code if necessary.

**Enforcement:** Enforcement was discussed exclusively by participants in the expert focus group, who emphasized its importance for users when using PPAs. However, they acknowledged the uncertainty about how enforcement is currently regulated and works in practice. As a potential mechanism for private enforcement, one participant suggested class action lawsuits, which could allow users to hold PPA providers accountable for non-compliance or privacy violations.

## Systemic conditions

**Public vs. market providers:** The expert focus group primarily preferred PPAs provided and run by public bodies. They worried that market PPAs could manipulate users by creating privacy preference filter bubbles, foster a privacy divide by commodifying privacy protection, and diminish public interests in how personal data circulates.

Participants in all focus groups expressed strong distrust toward big tech companies and, by extension, any PPA developed by them. Big tech companies were considered untrustworthy due to their history of non-compliance and privacy violations. Startups and small companies were dismissed by user focus groups as PPA providers due to their lack of scalability and the likelihood of big tech acquisitions, which diminished their PPAs' perceived neutrality. Likewise, NGOs as PPA providers were ruled out due to their lack of power and financial independence from their funders, compromising their neutrality and that of their PPAs. PPAs provided by public bodies like the EU were considered a more trustworthy and less infringing provider for PPAs due to their transparency, which was believed to come under higher levels of regulatory scrutiny and restrictions on data usage. At the same time, public PPAs were also cautioned with risks of violations regarding user self-determination, freedom of choice, and government paternalism by some participants from all focus groups. Specifically, fears of authoritarian states and weaponizing PPAs for surveillance were discussed on public PPAs.

Some participants ultimately advocated for a private-public model, such as a publicly funded PPA developed by a private company, which might be subject to higher standards of transparency and accountability than private PPAs. And, surprisingly, after all the discussion, some participants from user focus groups concluded that commercial PPAs developed by private companies would be the more *realistic* and *effective* option. They argued that private companies have the resources and incentives to build, maintain, and continuously improve their technologically advanced PPAs, compared to public PPAs. However, participants emphasized that such commercial PPAs need scrutiny under EU law and oversight.

**Free vs. fee:** The expert focus group was harmonized on PPAs as a free service to users. They raised the idea of PPAs as "techcare," drawing a parallel to public healthcare. The service was considered to be high-quality and freely accessible to all. If a PPA were free as a commercial service, it raised concerns about the hidden costs of a "free" commercial PPA. Users might pay in other ways, such as being subjected to advertisements or having their data collected and used for alternative revenue streams.

User focus groups, in contrast, preferred paid PPAs, which could offer a basic free version and a premium upgrade (e.g., removing ads or unlocking advanced features). Users can choose to stay in the free version. Participants also proposed different payment structures, including a one-off purchase and a subscription model with payment frequencies ranging from monthly to annual. The exact amount varied from a one-time payment of €50–100, an annual subscription of €40, to monthly fees ranging from €0.99 to €15. Nonetheless, willingness to pay depended on whether a PPA demonstrably provides greater privacy protection than non-use, a benefit that is inherently difficult to evaluate ex ante. Beyond direct consumer sales, participants suggested a business-to-business model in which firms acquire PPAs for employees to meet corporate privacy and security requirements, as well as a license model that allows one account to cover multiple devices.

Opponents of paid PPAs doubted the profiting from an individual's right to privacy and challenged the necessity of PPAs, arguing that privacy pop-ups may be annoying, but it does not justify paying for a service to deal with them. They noted that despite the inconvenience of making privacy decisions manually, they had never experienced any negative consequences from it, making PPAs seem more like a convenience than a necessity.

**Avoiding monopoly:** User focus groups expressed concerns about the centralization of personal information stored in a PPA, perceiving it as inherently risky and potentially leading to a monopoly of PPAs over user information. A PPA with exclusive control and access over user data could act against users' best interests, particularly if the goals of its developer misalign with those of users. Further concerns were the invasiveness of PPAs, potentially monitoring users' online activities, and overriding privacy decisions. Additionally, a single PPA and its algorithm led to concerns that it would dominate the landscape, branding dissenting privacy perspectives as illegitimate.

**Balancing risks and interests:** The expert focus group raised concerns that end-users may become overdependent on PPAs. Without limiting data collection and use, PPAs risk reproducing the very privacy harms they are envisioned to solve. The experts also cautioned about the "tragedy of the commons," related to the potential unintended consequences of individual privacy choices. The concerns were that the irrational privacy preferences of individual users could lead to suboptimal privacy decision-making at the individual level, and privacy preferences based solely on one's perceived best interests could harm the collective best interests at a group level. In addition, concerns were expressed about the potential manipulation of users by market PPAs, as well as the use of nudges by PPAs to sway user privacy preferences and decisions away from their best interests, and toward the interests of PPA providers.

## Discussion

Our findings partially resonate with the themes found in the pilot study of Stöver et al. (2023). The convergence on design-related elements demonstrates how important they are to the acceptance of PPAs and how some elements are shared by different cohorts of people. By providing more nuanced insights and other important themes integral to user acceptance, we extend the current understanding of factors influencing PPA acceptance and adoption. This section explores how focus group findings could contribute to acceptance models (specifically, UTAUT) for PPAs, offer user-centered implications for stakeholders of PPAs and AI assistants, and present a critical reflection on PPAs as a technology.

### Extending technology acceptance models on AI assistants for privacy management

The three qualitative themes identified in this study could be integrated into established models on technology acceptance to better explain and understand how users would adopt and employ PPAs, particularly the Unified Theory of Acceptance and Use of Technology (UTAUT). Below, we discuss how these themes can contribute to UTAUT, either by enriching contexts of existing constructs or by serving as theoretical extensions.

The theme of **Design Elements**, which includes perceived usefulness, mandatory features, information sources, and transparency, aligns closely with the Technology Acceptance Model (TAM). In TAM, perceived usefulness is a factor that influences the actual system use; while system design characteristics that contain the other sub-themes shape the factors on actual system use (Marangunić and Granić 2015). When Venkatesh et al. (2003) developed UTAUT, perceived usefulness from TAM was merged to a new construct named Performance Expectancy, defined as "the degree to which an individual believes that using the system will help him or her to attain gains in job performance." System design characteristics were not woven as an explicit construct in UTAUT. However, Venkatesh (2022) proposed technology characteristics like transparency as one of the viable directions for the adoption and use of AI tools. As a result, in the context of PPAs, mandatory features, information sources, and transparency could be modeled as contextual variables that strengthen users' performance expectancy on PPAs.

The theme of **External Conditions**, encompassing trust, regulations, interoperability, literacy education, oversight, and enforcement, could expand the existing UTAUT construct of Facilitating Conditions. Facilitating Conditions in

UTAUT refers specifically to technical and organizational resources necessary to support technology use (Venkatesh et al. 2003). Interoperability and literacy education are well in line with such a definition as they provide the technical and organizational support integral to facilitating PPA use. Because participants viewed regulatory and oversight mechanisms as equally important to technical and organizational resources for PPA adoption, there are two modeling choices for regulation and oversight: (a) broaden the facilitating conditions construct by adding items that capture the two sub-themes, or (b) create a separate construct that predicts behavioral intention alongside Facilitating Conditions. Trust has been considered important by many research on technology acceptance. A meta-analysis on UTAUT 2 finds that trust is one of the most added extensions to UTAUT (Tamilmani, Rana, and Dwivedi 2021). It is also placed under facilitating conditions (Alasmari 2024) or as a dimension of new constructs (Bajunaied, Hussin, and Kamarudin 2023), in addition to as an independent new construct (Al-Saedi et al. 2020; Khalilzadeh, Ozturk, and Bilgihan 2017). While findings from this article highlight the importance of trust on PPA acceptance, future research is needed to examine whether trust serves as a distinct construct of UTAUT for PPAs or not.

The influence of **Systemic Conditions** could create a distinct construct for UTAUT. Sub-themes in Systemic Conditions cannot be categorized under the original UTAUT constructs. And the theme captures systemic prerequisites that need to be satisfied before users have the behavioral intention to use the technology. Nonetheless, Systemic Conditions as a new theoretical construct for UTAUT demands rigorous construct validation before it can be integrated into the extended UTAUT model on PPAs.

## Implications of acceptability factors for stakeholders involved

Our focus group findings provide PPA stakeholders with a nuanced understanding of the expectations of prospective users and the insights of domain experts. The subsection below summarizes these insights into recommendations. Importantly, many of the insights voiced by participants and experts align with concerns long recognized in privacy and personalized privacy assistant scholarship (e.g., Colnago et al. 2020; Stöver et al. 2023), legal and ethics literature (e.g., Fairfield and Engel 2015 on the externalities of individual privacy maximization, Susser 2019 on individuals' lack of meaningful control), and AI research (e.g., Natale et al. 2020 on users' loss of control using AI voice assistants). This convergence demonstrates the shared concerns between researchers and users, the need to address these concerns, and the necessity of action.

**PPA developers and providers:** PPA developers should prioritize user-centered designs, emphasizing the usefulness of their assistants by providing time efficiency, minimal collection of personal information, and concrete transparency measures through features like open data or data logs on PPA decisions. To respect users' control and their privacy preferences, developers should encourage users to periodically reconsider whether the decisions made by PPAs align with their genuine privacy preferences. Building and maintaining user trust should also be central to the development process.

Specific mandatory features highlighted in this study, such as meaningful consent and withdrawal rights, and a reliable "kill switch" for PPA to execute data deletion requests, should be included. Previous research has advocated the installation of a kill switch in AI so that humans could temporarily or permanently stop the operation of an AI (Turner 2018). In contrast, users in our focus groups expected PPAs to be the kill switch, with the expectation that they would delete the information stored in different platforms or services. Developers should also avoid designs that unintentionally infringe on others' privacy or create harmful externalities, and maximize the security of stored user information.

Providers should prioritize users' best interests, refraining from using PPAs as tools to collect additional personal data for profit or secondary purposes. Using state-of-the-art technologies and providing regular updates will increase user acceptance. When offering PPAs commercially, providers could consider licensing or one-time purchase options, which are more attractive to users than subscription-based models.

**Legislators, oversight, and enforcement:** Legislators should consider establishing standardized default settings for PPAs to better align them with fundamental privacy and constitutional values. Clarifying accountability for potential harms caused by PPAs, either legislatively or through judicial clarification, is also essential prior to the availability of PPAs to the general public. Users should be explicitly empowered with clear pathways to legally challenge PPA developers or providers for any resulting harm or damage.

Compatibility with existing local regulations, such as the GDPR and the AI Act, should be addressed before PPAs become widely available, for example, in the EU. Regulatory authorities, including Data Protection Authorities and market oversight bodies, should actively oversee PPAs to ensure ongoing compliance with relevant laws.

Although PPAs are a specific branch of AI assistants dedicated to facilitating user privacy, the lessons drawn from their acceptance could inform the broader acceptance of AI assistants as a whole. The substantive themes and sub-themes identified in our focus group study on PPA acceptance are equally relevant to other AI assistants and should be weighed carefully by designers, policymakers, and commercial stakeholders. Nevertheless, we recognize that the nuances within these themes and sub-themes may differ depending on the particular goals, contexts, and functions of each assistant.

## Broader considerations

AI assistants epitomize frictionless design: by automating routine tasks and streamlining processes they boost productivity of humans (Doron et al. 2024; Marikyan et al. 2022). This pursuit of friction-free interactions now shapes most contemporary digital technologies, built to optimize and streamline human activities (Dunn and Cureton 2019; Harshitha, S, and Devapictahi 2025). The removal of frictions, and thus frictionlessness, has become the core design philosophy driving innovation in Silicon Valley (Popiel and Vasudevan 2024), with the goal to "produce technologies that automate a growing sphere of human activities while posing progressively fewer cognitive and aesthetic demands" (Kemper and Jankowski 2024, p. 294).

The very appeal of PPAs, as illustrated in the focus groups, lies in delivering a time-saving, effort-free experience for privacy management. Yet, the promise is not without concerns about the technology and its providers. Building on the findings, we want to invite readers to start a conversation on PPAs and AI assistants in general, on whether PPAs will only serve as quick fixes that provide immediate ease or relief to user privacy issues, or whether their relief has longer-term effects. Will PPAs mask deeper systemic issues that prevent individuals from having stronger control and agency over their privacy, or will they resolve them?

Frictionlessness needs a limit. A "fully" frictionless PPA might leave little time and space for users to reflect and grow on their privacy preferences and management. Some design frictions, thus, might be necessary for PPAs as they are shown to increase user understanding on the operation of the technology and encourage users to engage with it in a more mindful manner (Cox et al. 2016; Mejtoft, Hale, and Söderström 2019). In addition, as PPAs enable frictionless privacy management, this shift toward delegation could create a transition where the agency and control individuals have in an already narrow and limited privacy negotiation space will be further diminished, which cannot be addressed without a systemic change to open up negotiation between service providers and users, such as via a right to customization (Tamò-Larrieux et al. 2021).

A PPA also does little to change the lack of meaningful control for users. For example, from a legal perspective, Art. 6(1) of the GDPR sets out six legal bases for processing personal data, and consent is only one of them. Data processing can therefore still be lawful when it is necessary for the performance of a contract or justified by the legitimate interests of a controller or a third party, which PPAs cannot stop. Furthermore, the loss of individual agency with frictionless PPAs could be further exacerbated if the providers of PPAs have different interests from those of their users. In the focus groups, participants expressed considerable distrust and concerns toward private companies, especially big tech, as potential providers of PPAs. PPAs from big techs could reinforce the instrumentarian power they have over individuals. Instrumentarian power is the power that "aims to predict, modify, and monetize behavior through multifarious data-collection devices and the creation of a mentality that unhesitatingly commodifies human experience" (Risse 2023, p.162). If PPAs become an instrument of big tech companies with "malicious" intent, then such companies would be able to systematically monitor, measure, predict, and potentially influence user behaviors through data derived from user interactions with PPAs or information stored in PPAs. In such a worst-case scenario, the relief from privacy management would mask a deeper loss of individual autonomy and agency in privacy.

Even if big tech companies do not develop and provide PPAs, as long as surveillance capitalism (see Zuboff 2019) is still the primary revenue generation model for most companies, they will remain strongly incentivized to collect, process, and monetize user information, hindering PPAs rather than collaborating with them.

As a result, real progress in privacy protection and privacy management requires systemic, structural changes that technological solutions, like PPAs alone, cannot address, since they patch the symptoms, not solving the causes of our privacy issues. Without tackling root causes, PPAs might serve merely as a temporary, technological fix—a band-aid placed over a wound that never truly heals.

For AI assistants, participants' discussion about who should provide PPAs and our reflection on how "frictionless" PPAs might extend the instrumentarian power of dominant players serves as a cautionary tale, highlighting the need for systemic changes in regulation and revenue models. Only with such changes can we ensure AI assistants are truly of the people, by the people, and for the people.

**Limitations and future research agenda**

This study has two primary limitations. First, there is an imbalance in numbers between the expert and user focus groups. Second and related, the expert focus group was primarily composed of researchers from legal and social science backgrounds, limiting the diversity of perspectives represented. Incorporating additional stakeholders such as computer scientists, human-computer interaction experts, legislators, enforcement officials, big tech representatives, and PPA insiders would enrich the findings, providing broader expertise and more comprehensive insights into the acceptability of PPAs.

Since the focus groups in this study were conducted exclusively in the EU, the findings primarily reflect EU-specific perspectives, shaped by EU legislation and privacy perceptions. Future research could explore PPA acceptability in other regions or jurisdictions and compare it with the EU. Such comparative analyses could reveal region-specific themes and show broader patterns for PPAs and AI assistants in terms of their acceptance. Additionally, future research can build upon these themes to empirically test and refine an extended UTAUT model to assess the acceptance and adoption of PPAs.

## Conclusion

In this article, we describe findings from a focus group study on the acceptability of personalized privacy assistants. Thematic analysis identified three themes: Design Elements, External Conditions, and Systemic Conditions. The themes and sub-themes demonstrate that PPA acceptance by users and experts is influenced by nuanced aspects and cannot be reduced to AI technology or privacy alone: External conditions and systemic conditions also play a role in the acceptance of PPAs. As a result, these themes should be considered by technology acceptance models like UTAUT as they might be necessary to explain PPA acceptance. These themes should be taken into further consideration for the acceptance of AI assistants in general. Although users praised the design of PPAs for their ability to save time on privacy management, we reflect on PPAs as a frictionless technology, their potential facilitation of instrumentarian power, and their role as a technological band-aid to privacy issues.

# Appendix A

**Expert Focus Group**

What do you think are the three must-have, no-go, and nice-to-have features for TamagotchIA? Why?

**FG 2-3**

1. What is your initial impression of TamagotchIA? How do you perceive it might be useful?
2. Who is more acceptable for you to provide TamagotchIA? The EU? States? Big Techs like Google? NGOs? Or someone else? Why?
3. What are the must-have features TamagotchIA needs to have for you to accept it? Sub-question: What about the no-go features to avoid?
4. Imagine TamagotchIA can use various information sources to provide personalized privacy for you, such as your privacy preferences, previous privacy settings, online behavior, and demographics; which source of information do you feel most comfortable with and why?
   (a) Sub-question: Are there specific concerns or preferences you have regarding how TamagotchIA may use certain information to personalize privacy disclosure or decisions?
   (b) Sub-question: How important is it for you to know exactly what data it collects and uses to personalize privacy disclosure or decisions? Low, moderate or high and why?
5. How likely would you accept TamagotchIA once it becomes available, and in which aspect? Why or why not?

**FG 4-5**

Opening:
Can you briefly explain your main privacy concerns online? For example, in social media, web browsing?

1. What is your initial impression of TamagotchIA? Would it be useful? Please explain. To what degree do you expect the TamagotchIA will benefit your privacy? Provide examples of possible benefits.
2. Do you think TamagotchIA will be easy to use? If not, what efforts do you associate with its use?
3. Do you believe your family or friends should use TamagotchIA once it is available? Why?
4. What resources and support would you expect to be available in order to use TamagotchIA?
5. Would you enjoy using TamagotchIA? Why or why not?
6. Do you think people should pay for TamagotchIA? If so, how much do you think TamagotchIA should cost? Why?
7. Do you think that using TamagotchIA could become a habit? Do you see any potential barriers to using it in your daily digital activity?
8. Who is more acceptable for you to provide TamagotchIA? The EU? States? Big Techs like Google? NGOs? Or someone else? Why?

Closing: How likely would you accept TamagotchIA once it becomes available, and in which aspect? Why or why not?